\documentclass[lettersize,journal]{IEEEtran}
\usepackage{amsmath,amsfonts}
\usepackage{algorithmic}
\usepackage{algorithm}
\usepackage{array}
\usepackage{color}
\usepackage[caption=false,font=normalsize,labelfont=sf,textfont=sf]{subfig}
\usepackage{textcomp}
\usepackage{stfloats}
\usepackage{url}
\usepackage{verbatim}
\usepackage{graphicx}
\usepackage{amssymb}
\usepackage{bm}
\usepackage{cite}
\usepackage{booktabs}
\begin{document}

\title{Secure Directional Modulation with Movable Antenna Array Aided by RIS}

\author{Maolin Li, Jingdie Xin, Feng Shu, Xuehui Wang, Yongpeng Wu, Cunhua Pan
	\thanks{This work was supported in part by the National Natural Science Foundation of China under Grant U22A2002, and  by the Hainan Province Science and Technology Special Fund under Grant ZDYF2024GXJS292; in part by the Scientific Research Fund Project of Hainan University under Grant KYQD(ZR)-21008; in part by the Collaborative Innovation Center of Information Technology, Hainan University, under Grant XTCX2022XXC07; in part by the National Key Research and Development Program of China under Grant 2023YFF0612900.}
\thanks{Maolin Li and Jingdie Xin are with the School
	of Information and Communication Engineering, Hainan University, Haikou,
	570228, China (E-mail: limaolin0302@163.com, xjingdie@163.com).}
\thanks{Feng Shu is with the School of Information and Communication Engineering and Collaborative Innovation Center of Information Technology, Hainan University, Haikou 570228, China, and also with the School of Electronic and Optical Engineering, Nanjing University of Science and Technology, Nanjing 210094, China (E-mail: shufeng0101@163.com).}
\thanks{Xuehui Wang is with the School of Mathematics and Statistics, Hainan Normal University, Haikou 571158, China (E-mail: wangxuehui0503@163.com).}
\thanks{Yongpeng Wu  is with the Shanghai Key Laboratory of Navigation and Location Based Services, Shanghai Jiao Tong University, Minhang, Shanghai, 200240, China (e-mail:  yongpeng.wu2016@gmail.com).}
\thanks{Cunhua Pan is with the National Mobile Communications Research Laboratory, Southeast University, China (e-mail: cpan@seu.edu.cn).}}



\maketitle

\begin{abstract}
In this paper, to fully exploit the performance gains from moveable antennas (MAs) and reconfigurable intelligent surface (RIS), a RIS-aided directional modulation \textcolor{blue}{(DM)} network with movable antenna at base station (BS) is established. \textcolor{blue}{Based on the principle of DM,} a BS equipped with MAs transmits legitimate information to a single-antenna user (Bob) while exploiting artificial noise (AN) to degrade signal reception at the eavesdropper (Eve). The combination of AN and transmission beamforming vectors is modeled as joint beamforming vector (JBV) to achieve optimal power allocation. The objective is to maximize the achievable secrecy rate (SR) by optimizing MAs antenna position, phase shift matrix (PSM) of RIS, and JBV. The limited movable range (MR) and discrete candidate positions of the MAs at the BS are considered, which renders the optimization problem non-convex. \textcolor{blue}{To address these challenges, an optimization method under perfect channel state information (CSI) is firstly designed, in which the MAs antenna positions are obtained using compressive sensing (CS) technology, and JBV and PSM are iteratively optimized. Then, the design method and SR performance under imperfect CSI is investigated. The proposed algorithms have fewer iterations and lower complexity. Simulation results demonstrate that MAs outperform fixed-position antennas in SR performance when there is an adequately large MR available.}
\end{abstract}

	\section{Introduction}
\IEEEPARstart{W}{ith} the development of technologies such as reconfigurable intelligent surfaces (RIS) and multiple-input-multiple-output (MIMO), the security requirements for 6G wireless networks are becoming increasingly prominent~\cite{Zhi2021,Xiang2024,Zhang2024}. \textcolor{blue}{The integration of physical layer security (PLS) techniques such as directional modulation (DM) with RIS has emerged as a promising approach to enhance the security of modern communication systems~\cite{Chorti2022,Shu2021,Wei2021}. By leveraging RIS's unique capability to dynamically manipulate wireless propagation channels, this innovative combination offers novel solutions for PLS enhancement. This approach significantly improves both the security performance and reliability of communication systems while maintaining cost-effectiveness~\cite{Lin2023}, as it does not require substantial additional hardware investments.}

 \textcolor{blue}{The principle of directional modulation (DM) is to maintain the signal constellation at desired users (Bob) for correct demodulation by jointly optimizing the amplitude responses and phase modes, while distorting the signal constellation of potential eavesdroppers (Eve)~\cite{Dong2022, Wang2021, He2022}. Two secure multicast schemes for $K$-group systems were proposed in~\cite{Shu2018}, demonstrating superior secrecy rate (SR) performance with/without channel statistics. To address line-of-sight security challenges in DM, the integration with RIS has been extensively investigated. It is worth mentioning that the research in~\cite{Wang2022} shows that the position of RIS will affect the communication performance, and three low-complexity schemes are designed to achieve high rate performance. In~\cite{Shu2024}, for Eve working in full-duplex mode, a hybrid RIS was introduced to maximize SR and achieved higher security performance than passive RIS. RIS have emerged as a pivotal enabling technology for DM systems, owing to their cost-effectiveness and unique capability to dynamically reconfigure wireless propagation channels.}
 
  \textcolor{blue}{The above studies are based on the assumption of perfect channel state information (CSI), which may be difficult to implement in practical applications. Channel estimation errors and time-varying channel characteristics may have a significant impact on system performance. To alleviate impact caused by imperfect CSI, a robust beamforming framework was designed in \cite{Gao2024} to maximize the sum-rate and obtained a performance of 62.5\% compared to perfect CSI. The approach in~\cite{Yang2025} achieved enhanced system performance by successively maximizing the sum-rate lower bound using statistical CSI, which is superior to methods based exclusively on CSI estimation. It is worth noting that the study in \cite{Zheng2023} showed that even if CSI is not perfect, deploying RIS is more advantageous than adding base station (BS) antennas in the presence of hardware damage. Reference \cite{Pala2025} introduced a deep reinforcement learning framework for worst-case robust security design, demonstrating a 19.9\% improvement in secrecy capacity compared to conventional optimization techniques. Therefore, RIS deployment demonstrates performance enhancement and practical implementability even under imperfect CSI conditions.}

\textcolor{blue}{In addition to deploying RIS in wireless channels to enhance system degrees of freedom (DoFs), antenna mobility has also gained significant research attention. Existing studies on fixed antenna array (FPA) systems demonstrate performance degradation in complex propagation environments characterized by multi-path fading and co-channel interference, due to their inherent lack of spatial adaptability~\cite{Xiao2024,Shao2025,Zhu2024b}. Compared to FPAs, cable-driven movable antenna (MA) technology provides additional spatial diversity gains by enabling dynamic antenna placement, thus achieving improved system performance~\cite{Yang2024,Ma2024}. Reference~\cite{Zhu2024a} proposed a parallel greedy ascent algorithm for MAs position optimization, and demonstrated that expanded movable range (MR) in broadband systems yield progressive performance enhancement. In~\cite{Hu2024}, a scenario with multiple colluding Eves was considered, and the achievable SR was maximized through joint optimization of transmission beamforming and antenna position. In~\cite{Zhu2023}, MAs were used to achieve full array gain in the desired directions and zero steering in the undesired directions through antenna phase and position optimization, which relies on accurate angle of directions estimation. In~\cite{Zhang2025}, the authors proposed a joint optimization framework for the beamforming vector, RIS phase shift matrix, and MAs positions to maximize the achievable transmission rate.}

\textcolor{blue}{However, the majority of studies optimistically assume that the positions of MAs can be freely adjusted within a specified region, which might not align with realistic conditions. The authors in~\cite{Wu2023, Mei2024} discretized the MR of the transmitting antenna into sample points and demonstrated higher received power gain compared to FPA by strategically selecting MAs positions. The proposed methods have high computational complexity, and the safety performance potential of RIS needs to be further explored. In addition, compressive sensing (CS) technology has also demonstrated the potential to solve discrete optimization problems. As demonstrated in~\cite{Zhang2020}, the authors developed a RIS-assisted position modulation scheme by exploiting array sparsity, thereby significantly reducing system implementation complexity.}

\textcolor{blue}{To investigate security in MAs systems with limited MR and movable positions, this paper proposes a RIS-assisted secure DM network, where BS is equipped with MAs. The finite candidate positions of MAs lead to a highly non-convex optimization problem that typically incurs prohibitive computational complexity. To address this challenge, we develop a low-complexity approach based on CS and singular value decomposition (SVD), which iteratively optimizes the MAs positions, joint beamforming vector (JBV), and RIS phase shift matrix (PSM) to maximize SR. The key contributions of this paper are summarized as follows:
\begin{enumerate}
	\item We consider an RIS-assisted DM network equipped with MAs, where the MAs movement is discrete and constrained within a limited range, i.e., the positions can only be selected from a finite set of candidate locations. Then, joint optimization problem of maximizing SR through JBV, PSM, and MAs positions selection is formulated, where the JBV is modeled as a combination of the transmit beamforming vector and AN.
	\item To address this non-convex optimization problem, we propose low-complexity methods based on CS and SVD that are applicable to both perfect and imperfect CSI. Specifically, the CS technique is employed to select MAs positions, while the JBV and PSM are optimized iteratively exploiting SVD. The proposed methods demonstrate significantly reduced computational complexity with only marginal performance degradation compared to the optimal exhaustive search approach.
	\item Through comparative analysis with FPA systems, we reveal a fundamental performance trade-off between MA quantity and MR. Notably, when sufficient MR is available, a system with fewer MAs can achieve comparable performance to FPA systems equipped with more antennas, thereby reducing hardware requirements. Furthermore, the MA implementation yields substantial improvements, including 11 dBm power savings and a 13.04\% enhancement in SR performance.
\end{enumerate}}

The remaining part of the paper is as follows. In Sec. \ref{sec:2}, the system model and problem formulation are presented. In Sec. \ref{sec:3}, a CS-based position selection method for MAs corresponding to perfect CSI is proposed. Then, the method is improved to be applicable to imperfect CSI in Sec. \ref{sec:4}. The simulation results are provided in Sec. \ref{sec:5}, followed by the conclusion in Sec. \ref{sec:6}.

Notations: a, $\mathbf{a}$ and $\mathbf{A}$ denote scalars, vectors, and matrices, respectively. $|\ |$, $||\ ||_0$, $||\ ||$ and $||\ ||_{\infty}$ represent the operation of taking modulus, $l_0$-norm, $l_2$-norm, and $l_{\infty}$-norm, respectively. $[\ ]^T$ and $[\ ]^H$ are the transpose and  conjugate-transpose operations, respectively. $\cdot$ and $\otimes$ represent dot product and Kronecker product operations, respectively. $\mathbb{E}[\ ]$ express expectation. $\text{diag}(\ )$ denotes the diagonal operator. $[\ ]^+$ is a value greater than or equal to zero. $\mathbf{Q}(n,:)$ and $\mathbf{c}(n)$ represent the $n$-th row of the matrix $\mathbf{Q}$ and $n$-th element of the vector $\mathbf{c}$, respectively. $\Re\{x\}$ and $\Im\{x\}$ represent taking the real and imaginary parts of $x$, respectively. $\Pr\{\ \}$ denotes probability. $\angle t_u$ denotes the phase of $t_u$. The main notations are illustrated in Table I.
\begin{table}[th]
	\centering
	  \caption{List of Main Notations}
	\setlength{\tabcolsep}{0.01pt}
	\renewcommand{\arraystretch}{1.2}
	\begin{tabular}{c c}
		\toprule
		Notation & Description  \\ \hline
		$d$ & The minimum spacing between candidate positions  \\ 
		$N_a$ & The number of MAs  \\ \
		$N$ & The number of candidate positions within the MR  \\ 
		$N_x$ & The number of candidate positions along the $x$-axis  \\ 
		$N_z$ & The number of candidate positions along the $z$-axis  \\ 
		$M$ & The number of RIS units  \\ 
		$M_y$ & The number of RIS units along the $y$-axis  \\ 
		$M_z$ & The number of RIS units along the $z$-axis  \\ 
		$\mathbf{x}$ & Transmit signal  \\ 
		$\mathbf{w}$ & JBV  \\
		$\hat{\mathbf{w}}$ & The reweighed JBV  \\ 
		$\bm{\delta}$ & The reweighing vector \\ 
		$\mathbf{\Phi}$ & PSM in diagonal matrix form \\ 
		$\bm{\varpi}$ & PSM in vector form \\ 
		$\mathbf{q}_u$ & The aggregated channel under perfect CSI corresponding to Bob  \\ 
		$\mathbf{q}_e$ & The aggregated channel under perfect CSI corresponding to Eve  \\ 
		$\hat{ \mathbf{q}}_u$ & The aggregated channel under imperfect CSI corresponding to Bob  \\ 
		$\hat{ \mathbf{q}}_e$ & The aggregated channel under imperfect CSI corresponding to Eve  \\ 
		${ \bm{\epsilon}}_u$ & The channel estimation error corresponding to Bob  \\ 
		${ \bm{\epsilon}}_e$ & The channel estimation error corresponding to Eve  \\ 
		$\mathbf{p}$ & The position selection vector  \\ 
		$t_u$ & The magnitude of useful signal at Bob  \\ 
		$t_e$ & The magnitude of received signal at Eve  \\ 
		$\angle t_u$ & The phase of useful signal at Bob  \\ 
		$\angle t_e$ & The phase of received signal at Eve  \\ 
		$\varepsilon_1, \varepsilon_2$ & The auxiliary variable used for calculating $t_u$ \\ 
		$\hat{\varepsilon}_1, \hat{\varepsilon}_2$ & The auxiliary variable used for calculating $t_e$ \\ \hline
	\end{tabular}
\end{table}
\section{System Model and Problem Formulation}\label{sec:2}
\subsection{System Model}
\begin{figure}
	\centering
	\includegraphics[width=0.45\textwidth, trim = 2 20 1 2,clip]{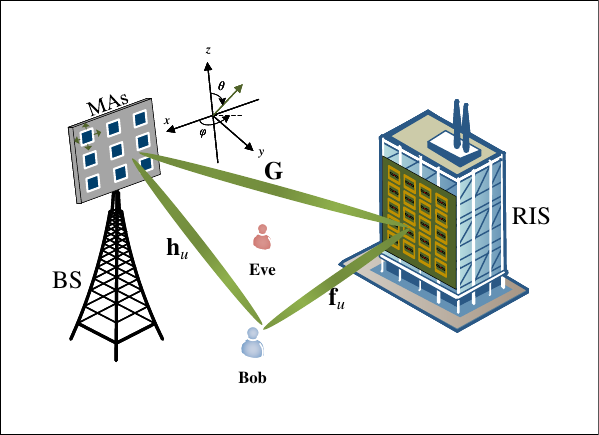}\\
	\caption{\textcolor{blue}{System for the proposed secure DM network with MA aided by active RIS.}}\label{fig:1}
\end{figure}

Fig. \ref{fig:1} illustrates an active RIS-assisted system, where the BS is equipped with $N_a$ MAs capable of moving within a range of $(N_x-1)d\times(N_z-1)d$, where $d$ is the minimum spacing between candidate positions. The electromagnetic units on the RIS are fixed in position and have a quantity of $M=M_yM_z$. The legitimate user (Bob) and eavesdropper (Eve) are equipped with a single antenna. Considering a three-dimensional space, where $N_xd$ and $N_zd$ represent the lengths in the x-axis and z-axis directions, and $M_y$ and $M_z$ represent the number of electromagnetic elements in the y-axis and z-axis directions, respectively. The MA array is positioned at the reference origin, with an actve RIS considered to be installed on the surface of the building, and Eve on the ground is closer to BS and RIS than Bob. \textcolor{blue}{Both the MA array and RIS are configured as two-dimensional planar arrays.} Defining $m\in\mathcal{M}=\{0, 1, \cdots, M-1\}$ and $n\in\mathcal{N}=\{0, 1, \cdots, N-1\}$, where $N$ represents the number of candidate positions within the MR.

The BS transmits legitimate information to Bob through the direct path and the path reflected by RIS. The elevation and azimuth angles of departure (AoD) from BS to RIS are represented as $\bar{\theta}$ and $\bar{\phi}$, respectively. Similarly, $\bar{\psi}$, $\tilde{\psi}$, $\hat{\phi}$, and $\tilde{\phi}$ denote the azimuth AoDs from the RIS to the Bob, from the RIS to the Eve, from the BS to the Bob, and from the BS to the Eve, respectively.  $\bar{\varphi}$, $\tilde{\varphi}$, $\hat{\theta}$, and $\tilde{\theta}$ denote the elevation AoDs corresponding to the transmission paths. $\phi$ and $\theta$ represent the azimuth and elevation angles of arrival (AoA) from the BS to the RIS. The transmit signal can be represented as
\begin{equation}
	\label{eq1}
	\textbf{x}=\mathbf{v}s+\mathbf{v}_a{z},
\end{equation}
where $s$ and $z$ are the transmission symbol with $\mathbb{E}[ss^H]=1$ and the AN with $\mathbb{E}[zz^H]=1$, respectively. $\mathbf{v}\in\mathbb{C}^{N\times1 }$ and $\mathbf{v}_a\in\mathbb{C}^{N\times1 }$ denote the beamforming vector of $s$ and $z$, respectively. We consider a JBV $\mathbf{w} = \mathbf{v}+\hat{\mathbf{v}}_a$, where $\hat{\mathbf{v}}_a = \mathbf{v}_azs^H$  is the null space vector of Bob, and then $\textbf{x}=\mathbf{w}s$, which can recover the original baseband signal through maximum ratio transmission  (MRT). \textcolor{blue}{The solution based on MRT for $\hat{\mathbf{v}}_a$ is illustrated in Appendix \ref{appendices}.}

To facilitate the antennas' movement, sampling the MR of the antenna with the minimum element spacing $d$ can obtain channels containing virtual and actual elements, where the virtual and actual elements are represented as antenna-free and MAs, respectively. The number of sampling points is $N=N_xN_z$. Then, the channels from the BS to the RIS, from the RIS to the Bob,  from the RIS to the Eve, from the BS to the Bob, and from the BS to the Eve are represented as $\mathbf{G}=(\mathbf{a}(\bar{\phi},\bar{\theta})\otimes\mathbf{b}(\bar{\phi},\bar{\theta}))\mathbf{g}^{T},
		\mathbf{f}_u=\bar{\mathbf{a}}(\bar{\psi},\bar{\varphi})\otimes\bar{\mathbf{b}}(\bar{\psi},\bar{\varphi}),
		\mathbf{f}_e=\bar{\mathbf{a}}(\tilde{\psi},\tilde{\varphi})\otimes\bar{\mathbf{b}}(\bar{\psi},\tilde{\varphi}),
		\mathbf{h}_u={\mathbf{a}}(\hat{\phi},\hat{\theta})\otimes{\mathbf{b}}(\hat{\phi},\hat{\theta})$, and
		$\mathbf{h}_e={\mathbf{a}}(\tilde{\phi},\tilde{\theta})\otimes{\mathbf{b}}(\tilde{\phi},\tilde{\theta})$, where $\mathbf{g}=\hat{\mathbf{a}}(\phi,\theta)\otimes\hat{\mathbf{b}}(\phi,\theta)$. $\mathbf{a}(\bar{\phi},\bar{\theta})\in\mathbb{C}^{N_x\times1 }$ and $\mathbf{b}(\bar{\phi},\bar{\theta})\in\mathbb{C}^{N_z\times1 }$ represent the steering vectors from the BS to the vertical and horizontal dimensions of the RIS, respectively, and can be written as
\begin{align}
				\mathbf{a}(\bar{\phi},\bar{\theta})&=\left[ 1,e^{j\frac{2\pi d\sin\bar{\phi}\sin\bar{\theta}}{\lambda}},
		...,e^{j\frac{2\pi(N_x-1)d\sin\bar{\phi}\sin\bar{\theta}}{\lambda}} \right] ^{T}, \label{eq2}\\
		\mathbf{b}(\bar{\phi},\bar{\theta})&= \left[1,e^{j\frac{2\pi d\cos\bar{\phi}}{\lambda}},...,e^{j\frac{2\pi (N_z-1)d\cos\bar{\phi}}{\lambda}}\right]^{T},\label{eq2-1}
\end{align}
where $\lambda$ is the wavelength, $N_x$ and $N_z$ represent the number of sampling points in the horizontal and vertical directions, respectively. Similarly,  $\bar{\mathbf{a}}(\bar{\psi},\bar{\varphi})\in\mathbb{C}^{M_y\times1 }$ and $\bar{\mathbf{b}}(\bar{\psi},\bar{\varphi})\in\mathbb{C}^{M_z\times1 }$,
$\bar{\mathbf{a}}(\tilde{\psi},\tilde{\varphi})\in\mathbb{C}^{M_y\times1 }$ and $\bar{\mathbf{b}}(\bar{\psi},\tilde{\varphi})\in\mathbb{C}^{M_z\times1 }$,
${\mathbf{a}}(\hat{\phi},\hat{\theta})\in\mathbb{C}^{N_x\times1 }$ and ${\mathbf{b}}(\hat{\phi},\hat{\theta})\in\mathbb{C}^{N_z\times1 }$, ${\mathbf{a}}(\tilde{\phi},\tilde{\theta})\in\mathbb{C}^{N_x\times1 }$ and ${\mathbf{b}}(\tilde{\phi},\tilde{\theta})\in\mathbb{C}^{N_z\times1 }$, and $\hat{\mathbf{a}}(\phi,\theta)^{N_x\times1 }$ and $\hat{\mathbf{b}}(\phi,\theta)^{N_z\times1 }$ represent the steering vectors of the vertical and horizontal dimensions of paths, respectively. Then, the signal received at Bob can be
expressed as
\begin{equation}
	\label{eq3}
	{y}=(\mathbf{f}_u^{H}\mathbf{\Phi}\mathbf{G}+\mathbf{h}^{T}_u)\mathbf{w}s+\mathbf{f}_u^{H}\mathbf{\Phi}\mathbf{n}_r+n_u,
\end{equation}
where $\mathbf{n}_r\in\mathbb{C}^{M\times1 }$ and $n_u$ represent additive Gaussian white noise (AWGN) with  distributions $\mathbf{n}_{r}\sim\mathcal{CN}\left(0,\sigma_{r}^{2}\mathbf{I}_M\right)$ and ${n}_{u}\sim\mathcal{CN}\left(0,\sigma_{u}^{2}\right)$ experienced by RIS and Bob, respectively. $\mathbf{\Phi}=\text{diag}(\bm{\varpi})\in\mathbb{C}^{M\times M}$ denotes PSM of RIS, and $\bm{\varpi}$ can be represented as
\begin{equation}
	\label{eq4}
\bm{\varpi} = [\alpha_{0}e^{j\vartheta_{0}},\alpha_{1}e^{j\vartheta_{1}},,\cdots,\alpha_{M-1}e^{j\vartheta_{M-1}}]^T,
\end{equation}
where \textcolor{blue}{$\alpha_m\in[0, \eta]$ and $\vartheta_m\in[0, 2\pi]$} are the adjustable phases and amplitudes of RIS, respectively. $\eta$ is the maximum amplification factor of RIS. The signal received at Eve can be written as
\begin{equation}
	\label{eq5}
	{e}=(\mathbf{f}_e^{H}\mathbf{\Phi}\mathbf{G}+\mathbf{h}_e^{T})\mathbf{w}s+\mathbf{f}_e^{H}\mathbf{\Phi}\mathbf{n}_r+n_e,
\end{equation}
where $n_e$ denote AWGN with variance $\sigma_{e}^{2}$ experienced by Eve.

Let $\mathbf{g}_u=\mathbf{f}_u^{H}\mathbf{\Phi}\mathbf{G}$ and $\mathbf{g}_e=\mathbf{f}_e^{H}\mathbf{\Phi}\mathbf{G}$ represent the cascaded channels reflected by RIS to Bob and Eve, respectively. $\mathbf{q}_u=\mathbf{g}_u+\mathbf{h}^{T}_u$ and $\mathbf{q}_e=\mathbf{g}_e+\mathbf{h}^{T}_e$ denote the aggregated channels of Bob and Eve, respectively. Then, the transmission rate of Bob can be represented as
\begin{equation}
	\begin{split}
		\label{eq6}
		r_u&=\log_2\left( 1+\frac{\mathbf{q}_u\mathbf{w}\mathbf{w}^H\mathbf{q}_u^H}{\sigma_r^2\mathbf{f}_u^{H}\mathbf{\Phi}\mathbf{\Phi}^H\mathbf{f}_u+\sigma_u^2}\right)=\log_2(1+\dfrac{t_u^2}{\kappa}), 
			\end{split}
	\end{equation}
where $t_u$ denotes the magnitude of useful signal at Bob, and $\kappa = \sigma_r^2\mathbf{f}_u^{H}\mathbf{\Phi}\mathbf{\Phi}^H\mathbf{f}_u+\sigma_u^2$ represents the sum of electromagnetic interference and noise power. The transmission rate of Eve can be represented as	
\begin{equation}
		\begin{split}
			\label{eq7}
		r_e&=\log_2\left(1+\frac{\mathbf{q}_e^H\mathbf{w}\mathbf{w}^H\mathbf{q}_e}{\sigma_r^2\mathbf{f}_e^{H}\mathbf{\Phi}\mathbf{\Phi}^H\mathbf{f}_e+\sigma_e^2}\right)=\log_2(1+\dfrac{t_e^2}{\varsigma}), 
	\end{split}
\end{equation}
where $t_e$ represents the magnitude of the signal corresponding to Eve, and $\varsigma=\sigma_r^2\mathbf{f}_e^{H}\mathbf{\Phi}\mathbf{\Phi}^H\mathbf{f}_e+\sigma_e^2$ is the sum of electromagnetic interference and noise power. Then, the SR of system is expressed as
\begin{equation}
	\begin{split}
		\label{eq8}
R_s&=\left[r_u-{r}_e\right]^+=\left[ \log_2\left( \frac{\kappa\varsigma+\varsigma t_u^2}{\kappa\varsigma+\kappa t_e^2}\right)\right]^+ .
	\end{split}
\end{equation}

\subsection{Problem Formulation}
 \textcolor{blue}{We aim to optimize JBV, PSM, and position selection of the MAs jointly to maximize the SR within the constraints of the maximum transmission power $P_0 $ and amplification factor $\eta$. To optimize the positions of the MAs, a position selection vector $\mathbf{p}=[p_0, p_1,\ldots, p_{N-1}]^T$ is defined, where $p_n\in\{0,1\}$ represents the binary selection variable for the $n$-th candidate position. Then, the optimization problem can be formulated as
\begin{equation}
	\begin{split}
		\label{eq9}
		\text{P1:}\quad\underset{\mathbf{w},\mathbf{\Phi},\mathbf{p}}{\max}&\ \ \ R_s
		\\ 
		\text{s.t.} &\quad\text{C1:}\ \mathbf{q}_u(\mathbf{w}\cdot\mathbf{p})=t_u,\\ 
		&\quad\text{C2:}\ \mathbf{q}_e(\mathbf{w}\cdot\mathbf{p})=t_es_es^H,\\
		&\quad\text{C3:}\ ||\mathbf{w}\cdot\mathbf{p}||\leq P_0,\\
		&\quad\text{C4:}\ \vartheta_{m}\leq\eta, m\in\mathcal{M},\\
		&\quad\text{C5:}\ {p}_n\in\{0,1\}, n\in\mathcal{N}.
	\end{split}
\end{equation}}
Here, constraint C1 represents the symbol design corresponding to Bob. $s_e$ denotes the rotation of legitimate information $s$, and constraint C2 represents the symbol synthesized at Eve. \textcolor{blue}{Constraints C3, C4, and C5 represent the threshold values for transmission power, the RIS amplification factor, and the binary constraint, respectively. }

\section{Proposed CS-based MAs Selection and SR Maximization Method with Perfect CSI}\label{sec:3}

In this section, we propose a CS-based method to solve the non-convex optimization problem P1 through alternating optimization (AO). 

\textcolor{blue}{In accordance with the principle of CS, the $l_0$-norm is exploited to select $N_a$ MAs, which can be denoted as $||\mathbf{w}||_0\leq N_a$. Thus, $\mathbf{p}$ can be determined by checking if the elements in $\mathbf{w}$ are zero. Then, P1 can be reformulated as
\begin{equation}
	\begin{split}
		\label{eq9-1}
		\text{P2:}\quad\underset{\mathbf{w},\mathbf{\Phi}}{\max}&\ \ \ R_s
		\\ 
		\text{s.t.} &\quad\text{C1}^{\prime}:\ \mathbf{q}_u\mathbf{w}=t_u,\\ 
		&\quad\text{C2}^{\prime}:\ \mathbf{q}_e\mathbf{w}=t_es_es^H,\\
		&\quad\text{C3}^{\prime}:\ ||\mathbf{w}||\leq P_0,\\
		&\quad\text{C4}:\ \vartheta_{m}\leq\eta, m\in\mathcal{M},\\
		&\quad\text{C5}^{\prime}:\ ||\mathbf{w}||_0\leq N_a.
	\end{split}
\end{equation}
The $l_0$-norm is non-convex, and P2 cannot be directly solved.}
We introduce the reweighting vector $\bm{\delta}=1/(|\mathbf{w}|+||\mathbf{w}||_{\infty})$ and auxiliary parameter $\xi$, approximate the $l_0$-norm with the $l_2$-norm, and constrain \textcolor{blue}{$\text{C5}^{\prime}$} is rewritten as
\begin{equation}
	\begin{split}
		\label{eq10}
\text{C5}^{\prime\prime}: ||\bm{\delta}\cdot\mathbf{w}|| \leq \xi.
	\end{split}
\end{equation}
Then, the sampling points corresponding to the first $N_a$ high values in $\mathbf{w}$ are considered as the positions of MAs. Let the SR be greater than the value $\mu$, i.e., $R_s\geq\mu$. According to \eqref{eq8}, we obtain
\begin{equation}
	\begin{split}
		\label{eq11}
		(t_u+\sqrt{\frac{2^\mu\kappa}{\varsigma}}t_e)(t_u-\sqrt{\frac{2^\mu\kappa}{\varsigma}}t_e)&=\rho(t_u-\tau t_e)\\&\geq(2^\mu-1)\kappa,
	\end{split}
\end{equation}
where $\rho=t_u+\tau t_e$ is the proportionality factor and $\tau=\sqrt{{2^\mu\kappa}/{\varsigma}}$. Then, the SR maximization problem can be reformulated as
\begin{equation}
	\begin{split}
		\label{eq12}
		\text{P3:}\quad\underset{\mathbf{w},\mathbf{\Phi}}{\max}&\ \ \ t_u-\tau t_e
		\\ 
		\text{s.t.} &\quad\text{C1}^\prime,\text{C2}^\prime,\text{C3}^\prime,\text{C4},\text{C5}^{\prime\prime}.
	\end{split}
\end{equation}
\subsection{Optimizing JBV and $t_u$ with fixed PSM}
Let the set of channels for Bob and Eve be represented as $\mathbf{Q}\in\mathbb{C}^{2\times N}$, where $\mathbf{Q}(1,:)=\mathbf{q}_u$ and  $\mathbf{Q}(2,:)=\mathbf{q}_e$. The set of transmission symbols for Bob and Eve is denoted as $\mathbf{c} = [s,s_e]^Ts^H$ and the set of magnitude is $\mathbf{t} = [t_u,t_e]^T$. Then, given the PSM of RIS, the sub-optimization problem can be expressed as
\begin{equation}
	\begin{split}
		\label{eq13}
		\text{P4:}\quad\underset{\mathbf{w}}{\max}&\ \ \ t_u-\tau t_e
		\\ 
		\text{s.t.} &\quad\text{C3}^\prime,\text{C5}^{\prime\prime}, \text{C6:}\  \mathbf{Q}\mathbf{w}=\mathbf{t}\cdot\mathbf{c}.
	\end{split}
\end{equation}
Problem P4 is convex and can be addressed by using the CVX toolbox. Nevertheless, it typically involves high complexity. We search for suboptimal solutions in the solution set space obtained through singular value decomposition (SVD).  

The solution set space of constraint C6 is represented as $\mathbb{S}$, and the span of the JVM can be written as $\text{span}(\mathbf{w})$. Then, we have
\begin{equation}
	\begin{split}
		\label{eq14}
\bm{\delta}\cdot\mathbf{w}\in\text{span}(\mathbf{w})\in\mathbb{S}.
	\end{split}
\end{equation}
According to \eqref{eq14} and constraint C6, let $\mathbf{w}=\bm{\delta}\cdot\hat{\mathbf{w}}$, we can obtain
\begin{equation}
	\begin{split}
		\label{eq15}
		&\mathbf{Q}({\bm{\delta}\cdot\hat{\mathbf{w}}})=([1\ 1]^T\bm{\delta}^T\cdot\mathbf{Q})\hat{\mathbf{w}}=\hat{\mathbf{Q}}\hat{\mathbf{w}}=\mathbf{t}\cdot\mathbf{c},
	\end{split}
\end{equation}
where $\hat{\mathbf{Q}}=[1\ 1]^T\bm{\delta}^T\cdot\mathbf{Q}$. $\hat{\mathbf{Q}}$ is not a full-rank matrix. Decomposing $\hat{\mathbf{Q}}$ using SVD can obtain $\hat{\mathbf{Q}}=\mathbf{USV}^H$. Then, 
\begin{equation}
	\begin{split}
		\label{eq16}
\hat{\mathbf{w}} = \mathbf{VT}^H\mathbf{U}^H(\mathbf{t}\cdot\mathbf{c})\in\mathbb{S},
	\end{split}
\end{equation}
 where $\mathbf{T}\in\mathbb{C}^{2\times N}$ represents the reciprocal of non-zero elements in $\mathbf{S}$. Considering that the constraint $\text{C3}^\prime$ equality holds, we have
\begin{equation}
	\begin{split}
		\label{eq17}
		||\hat{\mathbf{w}}||&=(\mathbf{VT}^H\mathbf{U}^H(\mathbf{t}\cdot\mathbf{c}))^H\mathbf{VT}^H\mathbf{U}^H(\mathbf{t}\cdot\mathbf{c})\\&=(\mathbf{t}\cdot\mathbf{c})^H\tilde{\mathbf{Q}}^H\tilde{\mathbf{Q}}(\mathbf{t}\cdot\mathbf{c})\\
		&=P_0,
	\end{split}
\end{equation}
where $\tilde{\mathbf{Q}}=\mathbf{VT}^H\mathbf{U}^H$.

To optimize $t_u$, we introduce the auxiliary variable $\varepsilon=t_u-t_e$, which denotes the objective function of P4, and substituting it into \eqref{eq17} to obtain the solution $\varepsilon_1$ and $\varepsilon_2$ of $\varepsilon$, i.e., 
\begin{align}
\varepsilon_1 &= \frac{-t_e(2a_0+b_0)+\sqrt{t_e^2(b_0^2-4a_0c_0)+4a_0P_0}}{2a_0},\label{eq18}\\
\varepsilon_2&= \frac{-t_e(2a_0+b_0)-\sqrt{t_e^2(b_0^2-4a_0c_0)+4a_0P_0}}{2a_0},\label{eq18-1}
\end{align}
where
\begin{align}
		a_0 &= (\tilde{\mathbf{Q}}(:,1) {\mathbf{c}}(1))^H\tilde{\mathbf{Q}}(:,1) {\mathbf{c}}(1),\label{eq19}\\
		b_0 &= 2\Re((\tilde{\mathbf{Q}}(:,1) {\mathbf{c}}(1))^H\tilde{\mathbf{Q}}(:,2) {\mathbf{c}}(2)),\label{eq19--1}\\
		c_0 &= (\tilde{\mathbf{Q}}(:,2) {\mathbf{c}}(2))^H\tilde{\mathbf{Q}}(:,2) {\mathbf{c}}(2)\label{eq19--2}.
\end{align}
According to \eqref{eq18} and \eqref{eq18-1}, $\varepsilon_1$ is greater than $\varepsilon_2$ and thus serves as the objective function value for P4. Given $t_e$, $t_u=t_e+\varepsilon_1$ can be obtained. Then, exploiting \eqref{eq10}, the top $N_a$ values in $|\hat{\mathbf{w}}|$ are selected as the weights for MAs. Accordingly, $\mathbf{w}$, $\mathbf{p}$, and $t_u$ can be obtained.
\subsection{Optimizing PSM and $t_e$ with fixed JBV}
To improve the MA system's robustness, we consider that even when the channel quality of the direct-path channel is degraded, the path reflected by RIS can be safely transmitted, satisfying
\begin{align}
\mathbf{g}_u\mathbf{w}&=\mathbf{f}_u^{H}\text{diag}(\mathbf{G}\mathbf{w})\bm{\varpi}=t_u,\label{eq20}\\
\mathbf{g}_e\mathbf{w}&=\mathbf{f}_e^{H}\text{diag}(\mathbf{G}\mathbf{w})\bm{\varpi}=t_es_es^H\label{eq20-1}. 
\end{align}
With \eqref{eq20}, \eqref{eq20-1} and constraints $\text{C1}^{\prime}$ and $\text{C2}^{\prime}$, we have $\mathbf{q}_u\mathbf{w}=t_u+\mathbf{h}_u^T\mathbf{w}$ and $\mathbf{q}_e\mathbf{w}=t_es_es^H+\mathbf{h}_e^T\mathbf{w}$. Let $\mathbf{Z}\in\mathbb{C}^{2\times N}$ represent the set of paths reflected by RIS, where $\mathbf{Z}(1,:)=\mathbf{f}_u^{H}\text{diag}(\mathbf{G}\mathbf{w})$ and  $\mathbf{Z}(2,:)=\mathbf{f}_e^{H}\text{diag}(\mathbf{G}\mathbf{w})$. Then, given the JBV of BS, the sub-optimization problem can be expressed as
\begin{equation}
	\begin{split}
		\label{eq21}
		\text{P5:}\quad\underset{\mathbf{\Phi}}{\max}&\ \ \ t_u-\tau t_e
		\\ 
		\text{s.t.} &\quad\text{C4}, \text{C7:}\ \mathbf{Z}\bm{\varpi}=\mathbf{t}\cdot\mathbf{c}.
	\end{split}
\end{equation}
\textcolor{blue}{Similarly, by performing SVD on $\mathbf{Z}$, a solution space for $\bm{\varpi}$ that satisfies constraint C7 can be derived, where ${\mathbf{Z}}=\mathbf{\hat{U}\hat{S}\hat{V}}^H$. Then, to satisfy constraint C4, we have
\begin{equation}
	\begin{split}
		\label{eq22}
		\bm{\varpi} = \frac{\eta\tilde{\mathbf{Z}}(\mathbf{t}\cdot\mathbf{c})}{|| \tilde{\mathbf{Z}}(\mathbf{t}\cdot\mathbf{c})||_{\infty}},
	\end{split}
\end{equation}	
which is obtained by normalization.} $\tilde{\mathbf{Z}}=\hat{\mathbf{V}}\hat{\mathbf{T}}^H\hat{\mathbf{U}}^H$, where $\hat{\mathbf{T}}\in\mathbb{C}^{2\times N}$ denotes the reciprocal of non-zero elements in $\hat{\mathbf{S}}$. 

\textcolor{blue}{Then, to calculate $t_e$, $\bm{\varpi}$ obtained from \eqref{eq22} is substituted into \eqref{eq20} and \eqref{eq20-1}, and the matrix $\mathbf{P}$ is constructed to represent the set of reflection channels, i.e.,
\begin{equation}
	\begin{split}
		\label{eq23}
		\mathbf{P}=\left[\begin{matrix}
			\mathbf{f}_u^{H}\text{diag}(\bm{\varpi})\mathbf{G}\\
			\mathbf{f}_e^{H}\text{diag}(\bm{\varpi})\mathbf{G}
		\end{matrix}\right].
	\end{split}
\end{equation}
Thus, we have $\mathbf{P}\mathbf{w}=\mathbf{t}\cdot\mathbf{c}$. According to the derivation in \eqref{eq17}, considering that the transmission power is $P_0$, we have}
\begin{equation}
	\begin{split}
		\label{eq24}
		(\mathbf{t}\cdot\mathbf{c})^H\tilde{\mathbf{P}}^H\tilde{\mathbf{P}}(\mathbf{t}\cdot\mathbf{c})=P_0,
	\end{split}
\end{equation}
where $\tilde{\mathbf{P}}=\tilde{\mathbf{V}}\tilde{\mathbf{T}}^H\tilde{\mathbf{U}}^H$ is obtained by decomposing $\mathbf{P}$ exploiting SVD. According to \eqref{eq24}, the two roots $\hat{\varepsilon}_1$ and $\hat{\varepsilon}_2$ of ${\varepsilon}$ can be obtained, i.e.,
\begin{align}
	\hat{\varepsilon}_1 &= \frac{t_u(2\hat{a}_0+\hat{b}_0)+\sqrt{t_u^2(\hat{b}_0^2-4\hat{a}_0\hat{c}_0)+4\hat{a}_0P_0}}{2\hat{a}_0}, \label{eq25}\\
	\hat{\varepsilon}_2&= \frac{t_u(2\hat{a}_0+\hat{b}_0)-\sqrt{t_u^2(\hat{b}_0^2-4\hat{a}_0\hat{c}_0)+4\hat{a}_0P_0}}{2\hat{a}_0} \label{eq25-1},
\end{align}
where
\begin{align}
		\hat{a}_0 &= (\tilde{\mathbf{P}}(:,2) {\mathbf{c}}(2))^H\tilde{\mathbf{P}}(:,2){\mathbf{c}}(2),		\label{eq26}\\
		\hat{b}_0 &= 2\Re((\tilde{\mathbf{P}}(:,1) {\mathbf{c}}(1))^H\tilde{\mathbf{P}}(:,2){\mathbf{c}}(2)),		\label{eq26-1}\\
		\hat{c}_0 &= (\tilde{\mathbf{P}}(:,1) {\mathbf{c}}(1))^H\tilde{\mathbf{P}}(:,1) {\mathbf{c}}(1)\label{eq26-2}.
\end{align}
\textcolor{blue}{Then, with $\hat{\varepsilon}_1\geq\hat{\varepsilon}_2$, $t_e=t_u-\hat{\varepsilon}_1$ can be obtained. The proposed method is summarized in Algorithm \ref{alg:alg2}. Asymptotically, the complexity of Algorithm \ref{alg:alg2} is dominated by the SVD operations, the max-value selection, and the number of iterations, resulting in an overall complexity of $\mathcal{O}(L(M^2+N^2+N\log_2N))$, where lower-order terms are omitted. $l$ represents the $l$-th iteration.}
\begin{algorithm}
	\caption{CS-based position selection method for MAs with perfect CSI.}\label{alg:alg2}
	\begin{algorithmic}[1]
		\STATE \textbf{Initialize}: Iteration index $l=0$, $P_0$, $\mathbf{\Phi}$, $t_e$
		\STATE\textbf{repeat}
		\STATE\hspace{0.5cm}$l=l+1$;
		\STATE\hspace{0.5cm}Decompose $\hat{\mathbf{Q}}$ using SVD;
		\STATE \hspace{0.5cm}Update $\hat{\mathbf{w}}$ according to \eqref{eq17};
		\STATE\hspace{0.5cm}Update $t_u=t_e+\varepsilon_1$ according to \eqref{eq18};
		\STATE\hspace{0.5cm}Select the top $N_a$ highest values in $|\hat{\mathbf{w}}|$ as the weights for MAs, and $\mathbf{p}$ can be obtained;
		\STATE\hspace{0.5cm}Decompose ${\mathbf{Z}}$ using SVD;
		\STATE\hspace{0.5cm}Update $\mathbf{\varpi}$ according to \eqref{eq22};
		\STATE\hspace{0.5cm}Update $t_e=t_u-\hat{\varepsilon}_1$ according to \eqref{eq25};
		\STATE\textbf{until} The objective function value converges or reaches the maximum number of iterations.
		\STATE Calculate $\mathbf{w}=\bm{\delta}\cdot\hat{\mathbf{w}}$.
		\STATE\textbf{Output}: $\mathbf{w}$ and $\bm{\varpi}$
	\end{algorithmic}
\end{algorithm}
\textcolor{blue}{\section{Proposed CS-based MAs Selection and SR Maximization Method with Imperfect CSI}\label{sec:4}
In this section, based on the derivation in Sec. \ref{sec:3}, an improved method based on CS is proposed to be applicable to imperfect CSI.  Building upon the constructive interference (CI) framework \cite{Wei2021}, we model both channel estimation errors and noise using Gaussian distributions. This probabilistic formulation permits the representation of received signals through statistical distributions, thereby achieving two key objectives: maintaining a high probability of correct symbol demodulation at Bob despite channel estimation inaccuracies and noise interference; and ensuring that symbols remain in the non-demodulation region with high probability at Eve.}
\textcolor{blue}{\subsection{Robust Model with Imperfect CSI}}	
\textcolor{blue}{Defining the channel estimation error vectors at Bob and Eve as $\bm{\epsilon}_u$ and $\bm{\epsilon}_e$, which can
be modeled by the Gaussian distributed variables as $\mathbb{CN}\{0,\sigma_{\epsilon,u}^2\mathbf{I}_N\}$ and $\mathbb{CN}\{0,\sigma_{\epsilon,e}^2\mathbf{I}_N\}$, respectively. According to \eqref{eq3} and \eqref{eq5}, the signals received at Bob and Eve can be denoted as
\begin{align}
	\label{eq28}
	\hat{y}&=\hat{\mathbf{q}}_u\mathbf{w}s+\mathbf{f}_u^{H}\mathbf{\Phi}\mathbf{n}_r+n_u,\\
	\hat{e}&=\hat{\mathbf{q}}_e\mathbf{w}s+\mathbf{f}_e^{H}\mathbf{\Phi}\mathbf{n}_r+n_e,
\end{align}
where $\hat{\mathbf{q}}_u = {\mathbf{q}}_u+\bm{\epsilon}_u$ and $\hat{\mathbf{q}}_e = {\mathbf{q}}_e+\bm{\epsilon}_e$ are the channels from BS to Bob and from BS to Eve corresponding to imperfect CSI, respectively. Then, the transmission rates of Bob and Eve can be represented as
\begin{align}
		\label{eq30}
		\hat{r}_u&=\log_2\left( 1+\frac{\hat{\mathbf{q}}_u\mathbf{w}\mathbf{w}^H\hat{\mathbf{q}}_u^H}{\sigma_r^2\mathbf{f}_u^{H}\mathbf{\Phi}\mathbf{\Phi}^H\mathbf{f}_u+\sigma_u^2}\right)=\log_2(1+\dfrac{t_u^2}{\hat{\kappa}}), \\
		\hat{r}_e&=\log_2\left(1+\frac{\hat{\mathbf{q}}_e^H\mathbf{w}\mathbf{w}^H\hat{\mathbf{q}}_e}{\sigma_r^2\mathbf{f}_e^{H}\mathbf{\Phi}\mathbf{\Phi}^H\mathbf{f}_e+\sigma_e^2}\right)=\log_2(1+\dfrac{t_e^2}{\hat{\varsigma}}), 
\end{align}
respectively. Then, SR of system can be expressed as
\begin{equation}
	\begin{split}
		\label{eq32}
		\hat{R}_s&=\left[\hat{r}_u-\hat{r}_e\right]^+=\left[ \log_2\left( \frac{\hat{\kappa}\hat{\varsigma}+\hat{\varsigma} t_u^2}{\hat{\kappa}\hat{\varsigma}+\hat{\kappa} t_e^2}\right)\right]^+ .
	\end{split}
\end{equation}
\textcolor{blue}{\subsection{Problem Reformulation}}	
Then, following a similar scheme to P2, the optimization problem can be rewritten as
\begin{equation}
	\begin{split}
		\text{P6:}\quad\underset{\mathbf{w},\mathbf{\Phi}}{\max}&\ \ \ \hat{R}_s
		\\ 
		\text{s.t.} &\quad\text{C1}^{\prime\prime}:\ \hat{\mathbf{q}}_u\mathbf{w}=t_u,\\ 
		&\quad\text{C2}^{\prime\prime}:\ \hat{\mathbf{q}}_e\mathbf{w}=t_es_es^H,\\
		&\quad\text{C3}^{\prime}:\ ||\mathbf{w}||\leq P_0,\\
		&\quad\text{C4}:\ \vartheta_{m}\leq\eta, m\in\mathcal{M},\\
		&\quad\text{C5}^{\prime\prime}:\ ||\bm{\delta}\cdot\mathbf{w}|| \leq \xi.
	\end{split}
\end{equation}
Due to imperfect CSI at both Bob and Eve, constraints $\text{C1}^{\prime\prime}$ and $\text{C2}^{\prime\prime}$ cannot always be satisfied, leading to signal deviation from ideal constellation points and consequently degraded SR performance. In \cite{Wei2021}, a DM design method based on statistical probability was proposed, which maximizes the signal power at Bob and places it in the demodulation region, while designing the phase at Eve to move it away from the demodulation region. Inspired by this, we rewrite the constraints $\text{C1}^{\prime\prime}$ and $\text{C2}^{\prime\prime}$ in the form of probabilities, i.e.,
\begin{align}
	\text{C8}:&\Pr\{|\Im\{(\hat{\mathbf{q}}_u\mathbf{w}s+\mathbf{f}_u^{H}\mathbf{\Phi}\mathbf{n}_r+n_u)s^H\}|\leq\notag\\&\quad(\Re\{(\hat{\mathbf{q}}_u\mathbf{w}s+\mathbf{f}_u^{H}\mathbf{\Phi}\mathbf{n}_r+n_u)s^H\}-\Delta t_u)\tan\varrho\}\geq\Gamma,\label{new6}\\
	\text{C9}:&\Pr\{|\Im\{(\hat{\mathbf{q}}_e\mathbf{w}s+\mathbf{f}_e^{H}\mathbf{\Phi}\mathbf{n}_r+n_e)s_e^H\}|\leq\notag\\&\quad(\Re\{(\hat{\mathbf{q}}_e\mathbf{w}s+\mathbf{f}_e^{H}\mathbf{\Phi}\mathbf{n}_r+n_e)s_e^H\})\tan\varrho\}\geq\Gamma,\label{new7}
\end{align}
where $\Gamma$ denotes the probability threshold of falling into the demodulation region. $\varrho = \tfrac{\pi}{B}$ represents the maximum phase deviation from the center of demodulation region, and $B$ denotes the constellation size. Physically, increasing $\Delta t_u$ enhances the received signal power at Bob, thus improving robustness against both imperfect CSI and noise. Substituting $\hat{\mathbf{q}}_u = {\mathbf{q}}_u+\bm{\epsilon}_u$ and $\hat{\mathbf{q}}_e = {\mathbf{q}}_e+\bm{\epsilon}_e$ into \eqref{new6} and \eqref{new7}, the constraints $\text{C6}$ and $\text{C7}$ can be rewritten as
\begin{align}
&	\text{C8}^{\prime}:\notag\\&\Pr\{|\Im\{(\mathbf{q}_u\mathbf{w}s+\bm{\epsilon}_u\mathbf{w}s+\mathbf{f}_u^{H}\mathbf{\Phi}\mathbf{n}_r+n_u)s^H\}|\leq\notag\\&(\Re\{(\mathbf{q}_u\mathbf{w}s+\bm{\epsilon}_u\mathbf{w}s+\mathbf{f}_u^{H}\mathbf{\Phi}\mathbf{n}_r+n_u)s^H\}-\Delta t_u)\tan\varrho\}\geq\Gamma,\label{8}\\
	&\text{C9}^{\prime}:\notag\\&\Pr\{|\Im\{({\mathbf{q}}_e\mathbf{w}s+\bm{\epsilon}_e\mathbf{w}s+\mathbf{f}_e^{H}\mathbf{\Phi}\mathbf{n}_r+n_e)s_e^H\}|\leq\notag\\
	&\quad(\Re\{({\mathbf{q}}_e\mathbf{w}s+\bm{\epsilon}_e\mathbf{w}s+\mathbf{f}_e^{H}\mathbf{\Phi}\mathbf{n}_r+n_e)s_e^H\})\tan\varrho\}\geq\Gamma.\label{9}
\end{align}
Since channel estimation error and noise follow Gaussian distribution with different variance, the linear combinations of $\Im\{\bm{\epsilon}_u\mathbf{w}\}$, $\Im\{\mathbf{f}_u^{H}\mathbf{\Phi}\mathbf{n}_rs^H\}$, $\Im\{n_us^H\}$, $\Re\{\bm{\epsilon}_u\mathbf{w}\}$, $\Re\{\mathbf{f}_u^{H}\mathbf{\Phi}\mathbf{n}_rs^H\}$, and $\Re\{n_us^H\}$ corresponding to Bob still follow Gaussian distribution with a modified variance. Thus, collecting all the uncertainty terms as variable $\hat{n}_u$, we have
\begin{align}
	&\text{C8}^{\prime\prime}:\notag\\&\Pr\{|\Im\{\mathbf{q}_u\mathbf{w}\}|\leq(\Re\{\mathbf{q}_u\mathbf{w}\}-\Delta t_u)\tan\varrho+\hat{n}_u\}\geq\Gamma,\label{38}
\end{align}
where $\hat{n}_u\sim\mathbb{CN}(0,\frac{N_a\sigma_{\epsilon,u}^2P_{0}+M\sigma_r^2\eta+\sigma_n^2}{2\cos^2\varrho})$. Similarly, collecting all the uncertain terms $\Im\{\bm{\epsilon}_e\mathbf{w}ss_e^H\}$, $\Im\{\mathbf{f}_e^{H}\mathbf{\Phi}\mathbf{n}_rs_e^H\}$, $\Im\{n_es_e^H\}$, $\Re\{\bm{\epsilon}_e\mathbf{w}ss_e^H\}$, $\Re\{\mathbf{f}_e^{H}\mathbf{\Phi}\mathbf{n}_rs_e^H\}$, and $\Re\{n_es_e^H\}$ corresponding to Eve as variable $\hat{n}_e$, we have
\begin{align}
	&\text{C9}^{\prime\prime}:\notag\\&\Pr\{|\Im\{\mathbf{q}_e\mathbf{w}ss_e^H\}|\leq(\Re\{\mathbf{q}_e\mathbf{w}ss_e^H\})\tan\varrho+\hat{n}_e\}\geq\Gamma.\label{39}
\end{align}
where $\hat{n}_e\sim\mathbb{CN}(0,\frac{N_a\sigma_{\epsilon,e}^2P_{0}+M\sigma_r^2\eta+\sigma_e^2}{2\cos^2\varrho})$. 
\begin{figure*}
\begin{align}
	\color{blue}\text{C8}^{\prime\prime\prime}:&\color{blue}\Pr\{\frac{(\Re\{\mathbf{q}_u\mathbf{w}\}-\Delta t_u)\tan\varrho-|\Im\{\mathbf{q}_u\mathbf{w}\}|}{\frac{\sqrt{N_a\sigma_{\epsilon,u}^2P_{0}+M\sigma_r^2\eta+\sigma_n^2}}{\sqrt{2\cos\varrho}}}\geq\frac{\hat{n}_u}{\frac{\sqrt{N_a\sigma_{\epsilon,u}^2P_{0}+M\sigma_r^2\eta+\sigma_n^2}}{\sqrt{2\cos\varrho}}}\}\geq\Gamma,\label{40}\\
	\color{blue}\text{C9}^{\prime\prime\prime}:&\color{blue}\Pr\{\frac{(\Re\{\mathbf{q}_e\mathbf{w}ss_e^H\})\tan\varrho-|\Im\{\mathbf{q}_e\mathbf{w}ss_e^H\}|}{\frac{\sqrt{N_a\sigma_{\epsilon,e}^2P_{0}+M\sigma_r^2\eta+\sigma_e^2}}{\sqrt{2\cos\theta}})}\geq\frac{\hat{n}_e}{\frac{\sqrt{N_a\sigma_{\epsilon,e}^2P_{0}+M\sigma_r^2\eta+\sigma_e^2}}{\sqrt{2\cos\varrho}}}\}\geq\Gamma,\label{41}
\end{align}
\hrulefill
\vspace*{4pt}
\end{figure*}
The inequalities in \eqref{38} and \eqref{39} can be seen as cumulative distribution function (cdf) of a Gaussian distributed variable. Normalizing the variances of $\hat{n}_u$ and $\hat{n}_e$ yields \eqref{40} and \eqref{41}, which are illustrated at the top of next page. Then, constrains $\text{C8}^{\prime\prime\prime}$ and $	\text{C9}^{\prime\prime\prime}$ can be denoted as
\begin{align}
	\text{C10}:&|\Im\{\mathbf{q}_u\mathbf{w}\}|\leq(\Re\{\mathbf{q}_u\mathbf{w}\}-\Delta t_u)\tan\varrho-\Theta_u,\label{50}\\
	\text{C11}:&|\Im\{\mathbf{q}_e\mathbf{w}ss_e^H\}|\leq(\Re\{\mathbf{q}_e\mathbf{w}ss_e^H\})\tan\varrho-\Theta_e,\label{51}
\end{align}
where $\Theta_u=\frac{\Phi^{-1}(\Gamma)\sqrt{N_a\sigma_{\epsilon,u}^2P_{0}+M\sigma_r^2\eta+\sigma_n^2}}{\sqrt{2}\cos\varrho}$ and $\Theta_e=\frac{\Phi^{-1}(\Gamma)\sqrt{N_a\sigma_{\epsilon,e}^2P_{0}+M\sigma_r^2\eta+\sigma_e^2}}{\sqrt{2}\cos\varrho}$. $\Phi^{-1}(\cdot)$ represents the inverse cdf of a standard Gaussian variable.}

\textcolor{blue}{Therefore, according to \eqref{eq11}, \eqref{50}, and \eqref{51}, the optimization problem under imperfect CSI can be reformulated as
\begin{equation}\label{44}
	\begin{split}
		\text{P7:}\quad\underset{\mathbf{w},\mathbf{\Phi}}{\max}&\ \ \ t_u-\tau t_e
		\\ 
		\text{s.t.} &\quad\text{C3}^\prime, \text{C4}, \text{C5}^{\prime\prime}, \text{C10}, \text{C11}.
	\end{split}
\end{equation}
To solve problem P7, the fixed-variable methods can be exploited. While the CVX toolbox~\cite{Research2012} provides a conventional solution method, its computational complexity is often high. To address this problem, we improve the iterative closed-form solution based on CS, which significantly reduces the computational overhead while maintaining solution accuracy.}

\textcolor{blue}{\subsection{Optimizing JBV and $t_u$ with fixed PSM}}

\textcolor{blue}{Define variables $\chi_u=\mathbf{q}_u\mathbf{w}=t_ue^{j\angle t_u}$ and $\chi_e=\mathbf{q}_e\mathbf{w}ss_e^H=t_ee^{j\angle t_e}$, which physically represent the received signals with phase rotation at Bob and Eve, respectively. Then, substituting $\chi_u$ and $\chi_e$ into \eqref{44}, we have
	\begin{equation}\label{45}
		\begin{split}
			\text{P8:}\quad\underset{\mathbf{w},\mathbf{\Phi}}{\max}&\ \ \ t_u-\tau t_e
			\\ 
			\text{s.t.} &\quad\text{C3}^\prime, \text{C4}, \text{C5}^{\prime\prime}, \\
			&\quad\text{C12}: |\Im\{\chi_u\}|\leq(\Re\{\chi_u\}-\Delta t_u)\tan\varrho-\Theta_u, \\
			&\quad\text{C13}: |\Im\{\chi_e\}|\leq(\Re\{\chi_e\})\tan\varrho-\Theta_e,\\
			&\quad\text{C14}: \mathbf{q}_u\mathbf{w}  = \chi_u,\\
			&\quad\text{C15}: \mathbf{q}_e\mathbf{w}ss_e^H = \chi_e.
		\end{split}
\end{equation}}

\textcolor{blue}{Given PSM, the sub-optimization problem with respect to JBV $\mathbf{w}$ can be represented as
\begin{equation}\label{54}
	\begin{split}
		\text{P9:}\quad\underset{\mathbf{w}}{\max}&\ \ \ t_u-\tau t_e
		\\ 
		\text{s.t.} &\quad\text{C3}^\prime, \text{C5}^{\prime\prime}, \text{C12}, \text{C13}, \text{C14}, \text{C15}.
	\end{split}
\end{equation}
Notably, in contrast to the strict alignment requirements of $\text{C1}^{\prime\prime}$ and $\text{C2}^{\prime\prime}$, constraints $\text{C14}$ and $\text{C15}$ permit controlled phase deviations between the transmitted signal and its ideal constellation point, with these deviations being governed by the boundary conditions specified in $\text{C12}$ and $\text{C13}$. Then, the approach is proposed to solve problem P9.}

\textcolor{blue}{According to the derivation of \eqref{eq18}, the phase of the rotation signal changes the value of $c_0$, while the values of $a_0$ and $c_0$ remain unchanged. We first compute the derivative of $\varepsilon_1$ with respect to $b_0$, yields
\begin{align}
	\frac{\partial \varepsilon_1}{\partial b_0}=\frac{-t_e}{2a_0}+\frac{t_e^2b_0}{2a_0\sqrt{t_e^2(b_0^2-4a_0c_0)+4a_0P_0}}.\label{55}
\end{align}
By calculating $\frac{\partial \varepsilon_1}{\partial b_0}=0$, we have
\begin{align}
	P_0 = t_e^2c_0.\label{56}
\end{align}}

\textcolor{blue}{Remark 1: According to the derivation in \eqref{56}, the monotonicity of $\varepsilon_1$ with respect to $b_0$ is related to the transmission power at BS, signal reception power at Eve, and the channel experienced by Eve. When $P_0 < t_e^2c_0$ with $b_0>0$, $\frac{\partial \varepsilon_1}{\partial b_0}>0$ and $\varepsilon_1$ monotonically increases with $b_0$, whereas when $P_0 \geq t_e^2c_0$ with $b_0\geq0$, $\frac{\partial \varepsilon_1}{\partial b_0}\leq0$ and $\varepsilon_1$ monotonically decreases with $b_0$. For all $b_0<0$, the inequality $\frac{\partial \varepsilon_1}{\partial b_0}<0$ holds, which guarantees that $\varepsilon_1$ is strictly monotonically decreasing. Evidently, according to \eqref{eq19--1}, the parameter $b_0$ increases monotonically with the phase difference $\nu=\angle t_u-\angle t_e$ ($0\leq\nu\leq\pi$) between the received signals at Bob and Eve, reaching its minimum value when the signals are in-phase ($\nu=0$) and attaining its maximum when they are out-of-phase ($\nu=\pi$).}

\textcolor{blue}{Nevertheless, the achievable value of $b_0$ is limited by constraints C12 and C13 and generally cannot reach the theoretical maximum. By applying Euler's formula, constraints C12 and C13 can be reformulated as
	\begin{align}
		&\text{C12}^{\prime}: |{t_u\sin\angle t_u}|\leq({t_u\cos\angle t_u}-\Delta t_u)\tan\varrho-\Theta_u, \label{57}\\
		&\text{C13}^{\prime}: |{t_e\sin\angle t_e}|\leq{t_e\cos\angle t_e}\tan\varrho-\Theta_e.\label{58}
\end{align}
According to Remark 1, given $P_0$, $t_e$, and $c_0$, $t_u$ is determined by $\nu$. Therefore, given a fixed $\angle t_u$, $\angle t_e$ can be derived from the monotonic relationship. In the case where $P_0 \geq t_e^2c_0$ and $b_0\geq0$ (with $t_u$ inversely proportional to $b_0$), maximizing $\nu$ under the constraint $\text{C13}^{\prime}$ yields the optimal $\angle t_e$. Specifically, the equation in \eqref{58} holds, we have
\begin{align}
 t_e^2(1+\tan^2\varrho)\cos^2\angle t_e&-2t_e\Theta_e\tan\varrho\cos\angle t_e\notag\\&+\Theta_e^2-t_e^2=0.\label{59}
\end{align}
By solving \eqref{59}, $\angle t_e$ is given by 
\begin{align} \angle t_e=\arccos(\tfrac{2t_e\Theta_e\tan\varrho\pm\sqrt{4 t_e^2(t_e^2(1+\tan^2\varrho)-\Theta_e^2)}}{2 t_e^2(1+\tan^2\varrho)}).\label{60}
\end{align}}

\textcolor{blue}{Similarly, the equation in \eqref{57} holds, we have
\begin{align}
	t_u^2(1+\tan^2\varrho)\cos^2\angle t_u&-2t_u(\Delta t_u\tan\varrho+\Theta_u)\tan\varrho\cos\angle t_u\notag\\&+(\Delta t_u\tan\varrho+\Theta_u)^2-t_u^2=0.\label{61}
	\end{align}
By solving \eqref{61}, $\angle t_u$ can be derived as \eqref{62}, as illustrated at the top of next page. 
\begin{figure*}[t]
\begin{align} 
	\color{blue}\angle t_u=\arccos(\frac{2t_u(\Delta t_u\tan\varrho+\Theta_u)\tan\varrho\pm\sqrt{4 t_u^2(t_u^2(1+\tan^2\varrho)-(\Delta t_u\tan\varrho+\Theta_u)^2)}}{2 t_u^2(1+\tan^2\varrho)}).\label{62}
\end{align}
	\hrulefill
\vspace*{4pt} 
\end{figure*}
According to \eqref{60} and \eqref{62}, the maximum value of $\nu$ that satisfies constraints C12 and C13 can be calculated. Then, given $t_u$ and $t_e$, $\mathbf{w}$ can be calculated by \eqref{eq16}, and $\mathbf{p}$ can be obtained by selecting the top $N_a$ values in $|{\mathbf{w}}|$. Consequently, JBV and $t_u$ can be derived based on PSM and $t_e$.}
\textcolor{blue}{\subsection{Optimizing PSM and $t_e$ with fixed JBV}}
\textcolor{blue}{Given JBV, the sub-optimization problem with respect to PSM $\mathbf{\Phi}$ can be represented as
	\begin{equation}\label{46}
		\begin{split}
			\text{P10:}\quad\underset{\mathbf{\Phi}}{\max}&\ \ \ t_u-\tau t_e
			\\ 
			\text{s.t.} &\quad\text{C4}, \text{C12}^{\prime}, \text{C13}^{\prime}, \text{C14}, \text{C15}.
		\end{split}
	\end{equation}
According to \eqref{eq22}, $\mathbf{\Phi}$ can be updated. To update $t_e$, the partial derivative of $\hat{\varepsilon}_1$ with respect to $\hat{b}_0$ can be given by
\begin{align}
	\frac{\partial \hat{\varepsilon}_1}{\partial \hat{b}_0}=\frac{t_u}{2\hat{a}_0}+\frac{t_u^2\hat{b}_0}{2\hat{a}_0\sqrt{t_u^2(\hat{b}_0^2-4\hat{a}_0\hat{c}_0)+4\hat{a}_0P_0}}.\label{64}
	\end{align}
	By calculating $\frac{\partial \hat{\varepsilon}_1}{\partial b_0}=0$, we have
\begin{align}
	P_0 = t_u^2\hat{c}_0.\label{65}
	\end{align}}	

\textcolor{blue}{Remark 2: According to the derivation in \eqref{65}, the monotonicity of $\hat{\varepsilon}_1$ with respect to $\hat{b}_0$ is related to the transmission power at BS, signal reception power at Bob, and the channel experienced by Bob. When $P_0 < t_u^2\hat{c}_0$ with $\hat{b}_0<0$, $\frac{\partial \hat{\varepsilon}_1}{\partial \hat{b}_0}<0$ and $\hat{\varepsilon}_1$ monotonically decreases with $\hat{b}_0$, whereas when $P_0 \geq t_e^2\hat{c}_0$ with $\hat{b}_0\leq0$, $\frac{\partial \hat{\varepsilon}_1}{\partial \hat{b}_0}\geq0$ and $\hat{\varepsilon}_1$ monotonically increases with $\hat{b}_0$. For all $b_0>0$, the inequality $\frac{\partial \varepsilon_1}{\partial b_0}>0$ holds, which guarantees that $\varepsilon_1$ is strictly monotonically increasing. According to \eqref{eq26-1}, the parameter $\hat{b}_0$ increases monotonically with the phase difference $\hat{\nu}=\angle \hat{b}_u-\angle\hat{b}_e$ ($0\leq\hat{\nu}\leq\pi$) between the received signals at Bob and Eve, reaching its minimum value when the signals are in-phase ($\hat{\nu}=0$) and attaining its maximum when they are out-of-phase ($\hat{\nu}=\pi$).}

\textcolor{blue}{Similarly, based on \eqref{60} and \eqref{62}, the optimal $\hat{\nu}$ can be computed. Through the optimization of PSM, channel reflected by RIS is accordingly enhanced. Then, leveraging \eqref{eq25}, parameter $t_e$ can be further optimized to achieve superior SR performance. Therefore, given JBV and $t_u$, PSM and $t_e$ can be optimized.}

\textcolor{blue}{The proposed method is summarized in Algorithm \ref{alg:alg3}. Asymptotically, the complexity of Algorithm \ref{alg:alg3} is dominated by the SVD operations, the max-value selection, and the number of iterations, resulting in an overall complexity of $\mathcal{O}(L(M^2+N^2+N\log_2N))$, where lower-order terms are omitted. $\hat{l}$ denotes the $\hat{l}$-th iteration.}
\begin{algorithm}
\caption{\color{blue}CS-based position selection method for MAs with imperfect CSI.}\label{alg:alg3}
	\begin{algorithmic}[1]
		\color{blue}\STATE \textbf{Initialize}: Iteration index $\hat{l}=0$,  $P_0$, $\mathbf{\Phi}$, $t_e$
		\STATE\textbf{repeat}
		\STATE\hspace{0.5cm}$\hat{l}=\hat{l}+1$;
		\STATE\hspace{0.5cm}Calculate $\angle t_e$ and $\angle t_u$ according to \eqref{60} and \eqref{62};
		\STATE \hspace{0.5cm}Given $P_0$, $t_e$, $\angle t_e$, and $\angle t_u$, compute $t_u$ using the derivations in Remark 1 and \eqref{eq18};
		\STATE\hspace{0.5cm}Update $\mathbf{w}$ according to \eqref{eq16}, where $\mathbf{w}=\bm{\delta}\cdot\hat{\mathbf{w}}$;
		\STATE\hspace{0.5cm}Select the top $N_a$ highest values in $|{\mathbf{w}}|$ as the weights for MAs, and $\mathbf{p}$ can be obtained;
		\STATE\hspace{0.5cm}Calculate $\angle t_e$ and $\angle t_u$ according to \eqref{60} and \eqref{62};
		\STATE \hspace{0.5cm}Given $P_0$, $t_u$, $\angle t_e$, and $\angle t_u$, compute $t_e$ using the derivations in Remark 2 and \eqref{eq25};
		\STATE \hspace{0.5cm}Update $\mathbf{\Phi}$ according to \eqref{eq22};
		\STATE\textbf{until} The objective function value converges or reaches the maximum number of iterations.
		\STATE\textbf{Output}: $\mathbf{w}$ and $\bm{\Phi}$
	\end{algorithmic}
\end{algorithm}

\section{Simulation Results}\label{sec:5}
In this section, simulation results are presented. In three-dimensional space, the positions of BS, RIS, Bob, and Eve located in the far field are [0, 0, 0], [-100, 1000, -10], [500, 1200, -50], and [400, 900, -50] in meters, respectively. Unless otherwise specified, the carrier frequency is set to 25 GHz, and the minimum spacing $d$ between MAs is $\lambda/2$. The noise power experienced by RIS, Bob, and Eve is $-80$ dBm. \textcolor{blue}{The MR of MAs includes $4d\times4d$, $8d\times8d$, $9d\times9d$, and $11d\times11d$, respectively. We compare the proposed CS-based MA selection method with three baseline schemes: 1) ``FPA arranged side by side”: The BS is equipped with a FPA consisting of $N_a$ elements, where adjacent antennas are spaced half a wavelength apart. 2) ``FPA with random position”: The BS is equipped with $N_a$ antennas, where the antennas are randomly selected within a range of $(N_x-1)d\times  (N_z-1)d$; 3) ``FPA with worst position”: The worst-performing antenna position combinations from a large number of samples is select, with each sample comprising $N_a$ elements. To ensure comprehensive performance evaluation, there are two trade-off schemes for comparative analysis: 1) ``Scheme 1'' represents the MA system with a fixed MR of $0.144m\times0.144m$. Compared to large-size antennas, small-size antennas provide more candidate positions due to smaller minimum antenna spacing. 2) ``Scheme 2'' denotes the MA system with a fixed number of candidate positions, i.e., 64 in our configuration. In contrast to small-size antennas, large-size antennas achieve a larger MR due to larger minimum antenna spacing. Note that the minimum distance between adjacent antennas is measured through the center of the unit.}
\begin{figure}
	\centering
	\includegraphics[width=0.45\textwidth, trim = 2 5 1 10,clip]{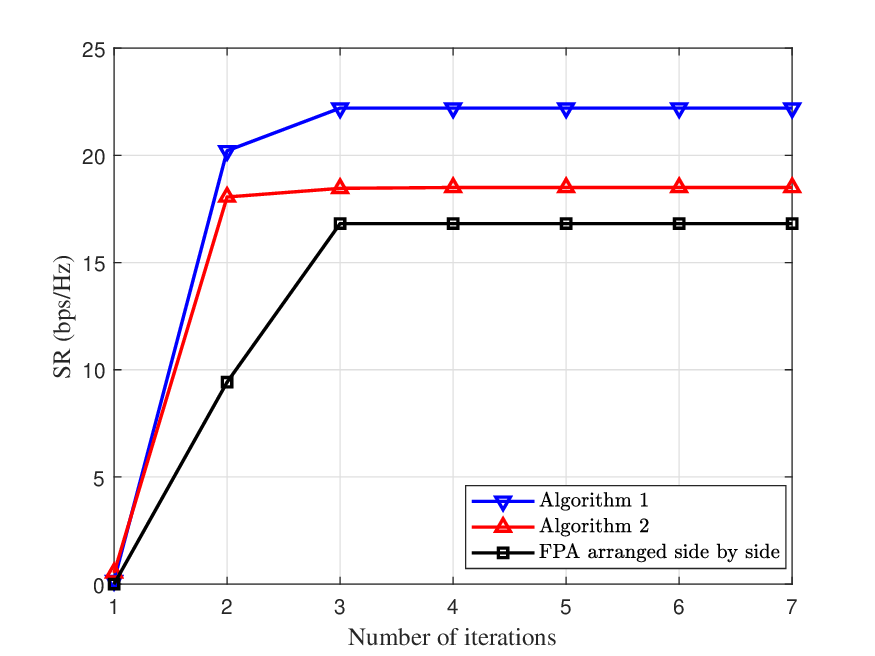}\\
	\caption{\textcolor{blue}{Convergence behaviour of the proposed algorithms.}}\label{fig:2_0}
\end{figure}
\begin{figure}
	\centering
	\includegraphics[width=0.45\textwidth, trim = 2 5 1 10,clip]{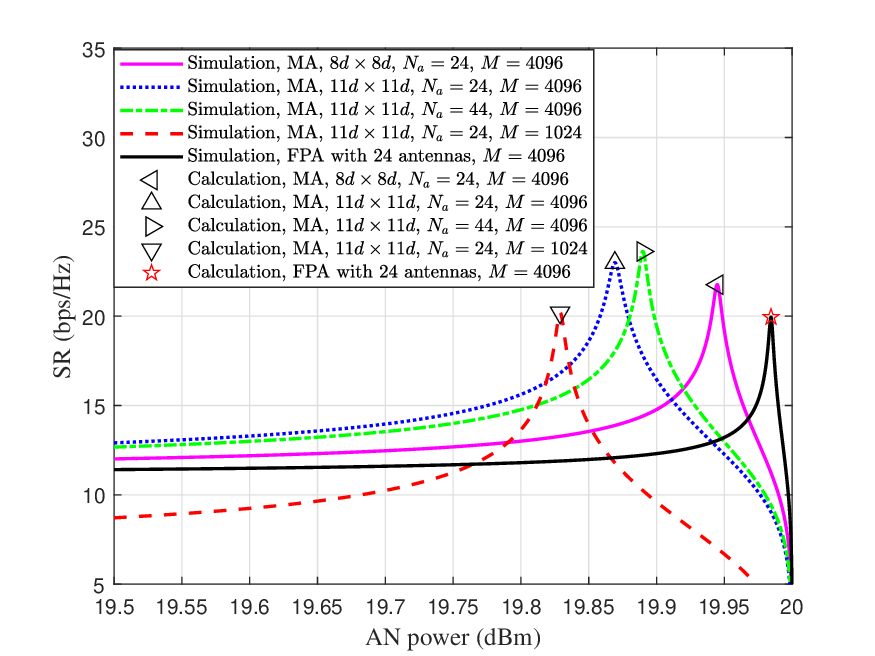}\\
	\caption{{\textcolor{blue}{SR versus AN power with $P_0=20~dBm$ and $\eta = 1.5$.}}}\label{1}
\end{figure}
\begin{figure}[t]
	\centering
	\includegraphics[width=0.45\textwidth, trim = 2 5 1 10,clip]{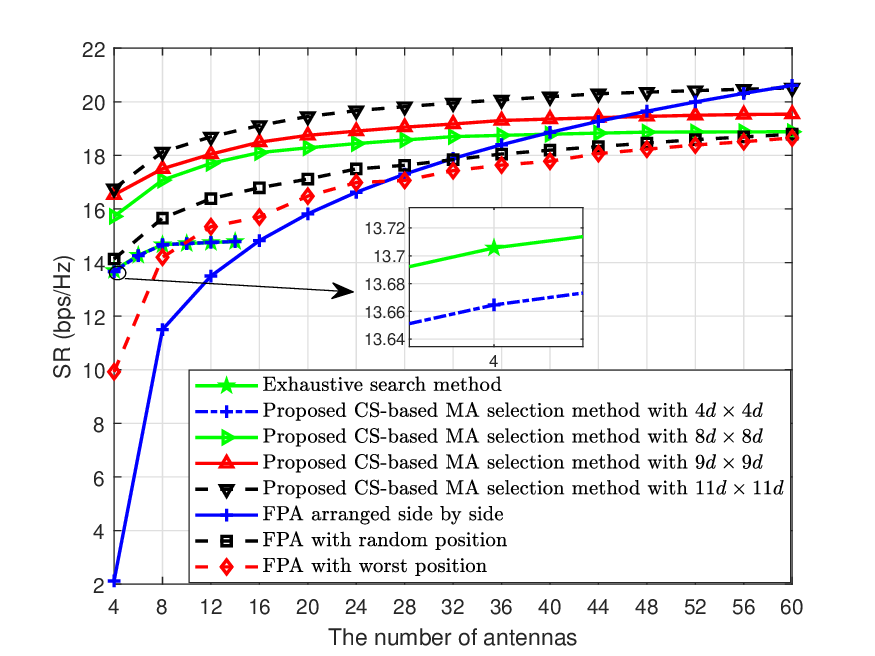}\\
	\caption{\textcolor{blue}{SR versus $N_a$ with $M=64\times64$, $\eta = 1.5$, and $P_0 = 20$ dBm.}}\label{fig:2}
\end{figure}
\begin{figure}
	\centering
	\includegraphics[width=0.45\textwidth, trim = 2 5 1 3,clip]{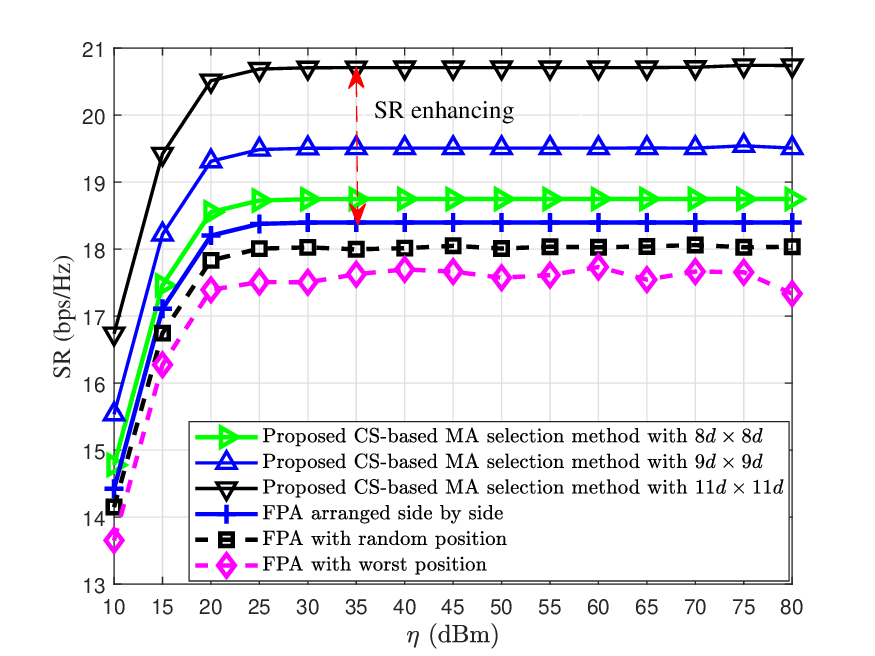}\\
	\caption{\textcolor{blue}{SR versus $\eta$ with $M=64\times64$, $N_a = 36$ and $P_0 = 20$ dBm.}}\label{fig:3}
\end{figure}

\textcolor{blue}{The convergence behaviors are investigated, as shown in Fig. \ref{fig:2_0}. The proposed method achieves convergence in fewer iterations, exhibiting lower complexity.} \textcolor{blue}{As illustrated in Fig. \ref{1}, the simulation results depict the SR under varying levels of AN power, with the calculation results derived from \eqref{30}. It is evident that the calculation results match the optimal SR obtained through simulation. Compared with FPA, MA can achieve higher SR and require lower AN power, demonstrating higher energy efficiency.}

Fig. \ref{fig:2} illustrates the SR under different number of antennas. The transmission power of BS and the amplification factor of RIS are set to $20$ dBm and $1.5$, respectively. \textcolor{blue}{As shown in Fig.  \ref{fig:2}, the FPA can achieve performance comparable to that of the MA when a sufficient number of antennas is employed. This is because increasing the number of antennas provides additional DoFs for the system. However, the additional selectable antenna positions in the MA sysyem offer more DoFs, resulting in superior performance. For instance, the FPA system equipped with 40 antennas achieves an SR of 18.8 bps/Hz, whereas the MA system with a MR of $9d\times9d$ requires only 24 antennas, saving 15\% in antenna count. As demonstrated by the SR values under different MRs, expanding the MR increases the candidate positions for MAs, leading to further SR performance improvements. The trade-off is that even with fewer antennas, the array must support a larger MR to ensure additional DoFs for the system. With a small number of antennas, equipping MAs can significantly enhance performance. For example, the MA system with four antennas and the MR of $8d\times8d$ achieves SR approximately eight times higher than that of the traditional side-by-side FPA system. Moreover, the SR level of MA system surpasses that of FPAs with random and worst-case positions, validating the effectiveness of proposed methods.} \textcolor{blue}{When the number of antennas is 4, the performance obtained by the CS-based position selection method only decreases by $0.094\%$ compared to exhaustive search, which proves the effectiveness of proposed method.} 

\begin{figure}[t]
	\centering
	\includegraphics[width=0.45\textwidth, trim = 2 5 1 1,clip]{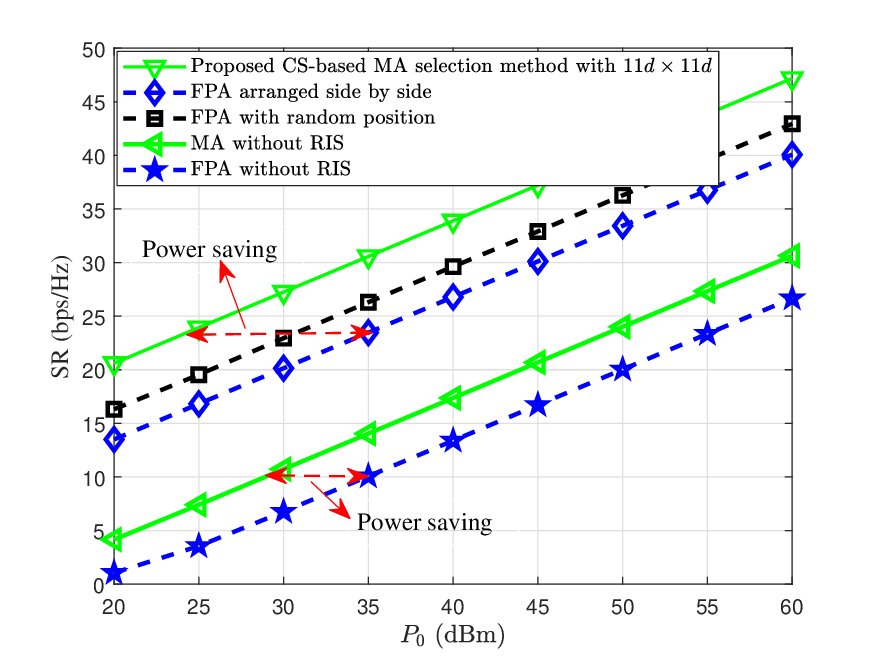}\\
	\caption{\textcolor{blue}{SR versus $P_0$ with $M=64\times64$, $N_a = 12$, and $\eta = 1.5$.}}\label{fig:4}
\end{figure}

\begin{figure}[t]
	\centering
	\includegraphics[width=0.45\textwidth, trim = 2 5 1 1,clip]{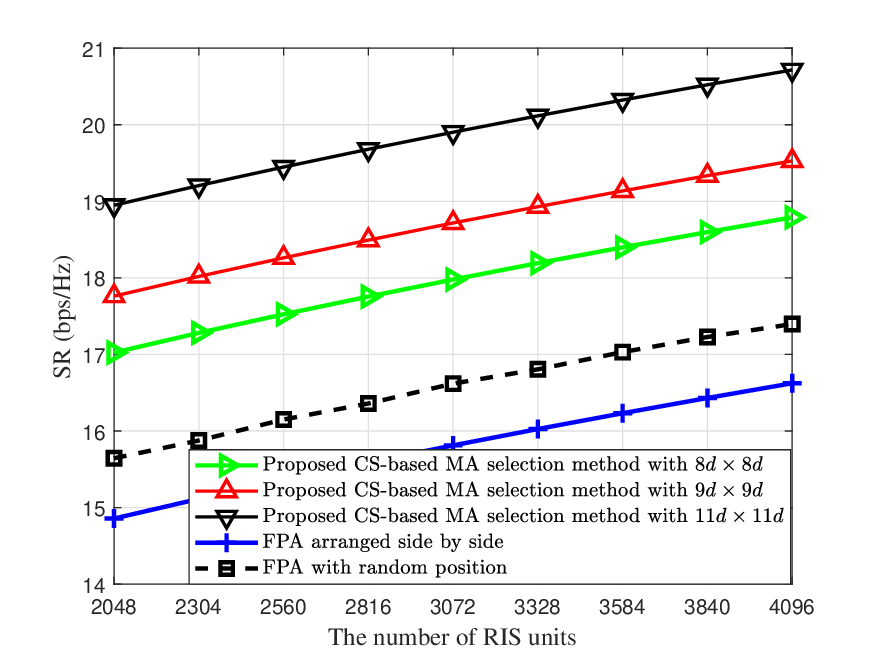}\\
	\caption{\textcolor{blue}{SR versus $M$ with $P_0 = 20$ dBm, $N_a = 24$, and $\eta = 1.5$.}}\label{fig:5}
\end{figure}

Fig. \ref{fig:3} illustrates the SR for different amplification factors of RIS, where $M=64\times64$, $N_a = 36$, and $P_0 = 20$ dBm. The SR values after the amplification factor of 25 dBm demonstrate a stable tendency. \textcolor{blue}{When the MRs are $8d\times8d$, $9d\times9d$, and $11d\times11d$, the corresponding SR enhancements are $1.9\%$, $6.0\%$ and $12.5\%$, respectively. Enlarging the MR can attain higher SR at the same $\eta$.} Fig. \ref{fig:4} describes the SR for different transmission powers, where $M=64\times64$, $N_a = 12$, and $\eta = 1.5$. \textcolor{blue}{``MA without RIS" and ``FPA without RIS" denote the MA and FPA systems operating without RIS assistance, respectively. In the absence of RIS, the MA achieves a 7 dBm power saving compared to the FPA. When RIS is deployed, the power saving further increases to 11 dBm. Clearly, the RIS-assisted DM system conserves significantly more transmit power compared to systems without RIS.} Fig. \ref{fig:5} illustrates the SR for different number of RIS units. \textcolor{blue}{The results indicate that increasing the number of RIS elements enhances the SR performance of MA systems.}

\begin{figure}[t]
	\centering
	\includegraphics[width=0.45\textwidth, trim = 2 1 1 10,clip]{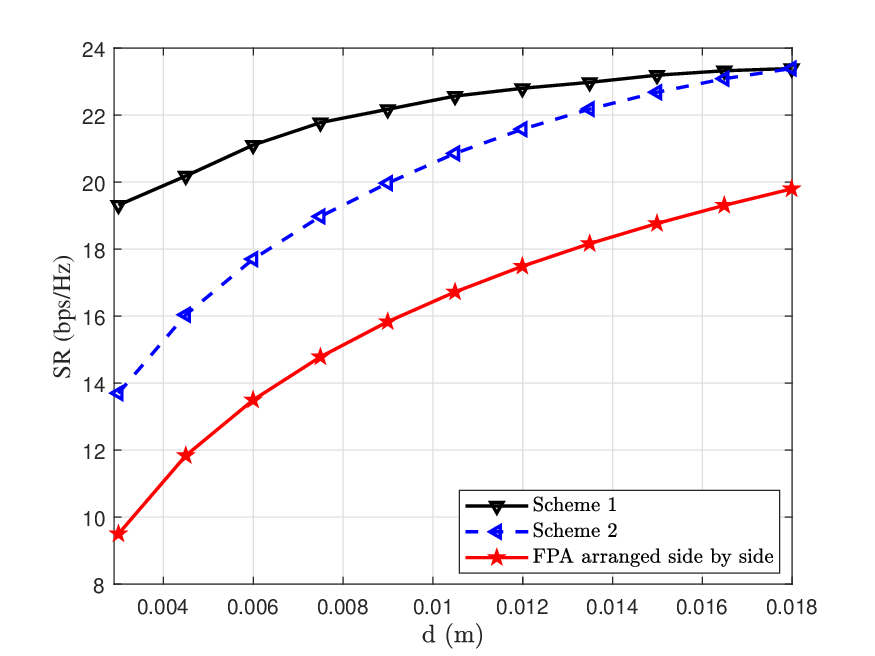}\\
	\caption{\textcolor{blue}{SR versus $d$ with $M=64\times64$, $\eta = 1.5$, and $P_0 = 20$ dBm.}}\label{fig:8}
\end{figure}

\textcolor{blue}{Fig. \ref{fig:8} illustrates the SR versus the minimum spacing $d$ between candidate positions. It can be observed that the larger-aperture antenna array can achieve a higher SR. For ``Scheme 1'', as $d$ increases, although the number of candidate positions decreases, the SR enhances. This indicates that increasing the antenna spacing is more critical for SR performance than increasing the number of candidate positions. For ``Scheme 2'', as $d$ increases, while the number of candidate positions remains unchanged, the antenna aperture and MR expand, leading to an enhanced SR. This further validates the advantage of larger-aperture antenna arrays in improving SR performance. Moreover, under the same parameter $d$, ``Scheme 1'' incorporates more candidate positions than ``Scheme 2'', providing additional DoFs for the MA system, thereby enhancing the SR.}

\begin{figure}[t]
	\centering
	\includegraphics[width=0.45\textwidth, trim = 2 1 1 10,clip]{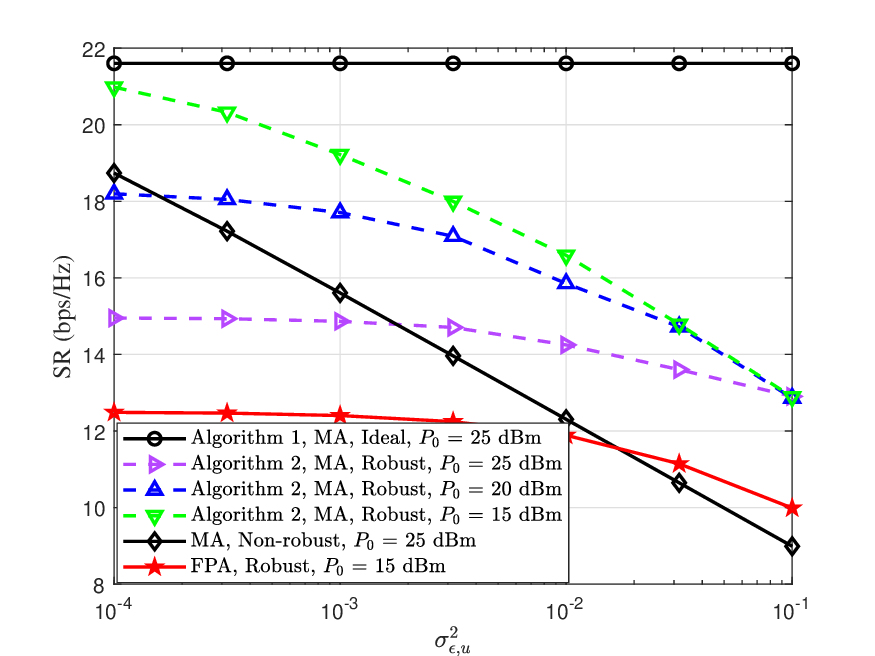}\\
	\caption{\textcolor{blue}{SR versus $\sigma_{\epsilon,u}^2$ with $M=64\times64$, $\eta = 1.5$, and $N_a = 20$.}}\label{fig:9}
\end{figure}

\textcolor{blue}{Fig. \ref{fig:9} illustrates the SR performance under different channel estimation error levels. With relatively small estimation errors (e.g., $\sigma_{\epsilon,u}^2=10^{-4}$), increasing the transmit power by 5 dBm can improve the SR by approximately 3 bps/Hz, approaching the ideal channel condition performance obtained by Algorithm \ref{alg:alg2}. However, as the channel estimation error increases, the SR improvement from higher transmit power gradually diminishes. Compared to FPA, the MA systems can achieve an SR gain of about 2.5 bps/Hz. Furthermore, compared to non-robust designs, the proposed Algorithm \ref{alg:alg3} achieves SR gain of approximately 2.2 bps/Hz. Therefore, in practical implementation, compared to FPA, the deployment of MA can achieve higher levels of secure communication, with good system robustness and advantages in cost savings.}


\section{Conclusions}\label{sec:6}
\textcolor{blue}{In this paper, an active RIS-aided secure DM network with MA was proposed and the secrecy performance was investigated. We aimed to maximize the secrecy rate through joint optimization of MAs' positions, JBV, and PSM. To address this challenge, two iterative optimization algorithms based on CS and SVD of low complexity were proposed for perfect and imperfect CSI scenarios, respectively. Specifically, the MA positions with finite candidate locations were optimized using CS, while JBV and PSM were optimized through SVD. Simulation results demonstrated the effectiveness of the proposed methods. Compared with FPA systems, the MA-based configurations can achieve superior secrecy performance and have advantages in both cost efficiency and system robustness.}
\textcolor{blue}{\begin{appendices}
	\section{Solution to $\hat{\mathbf{v}}_a$}\label{appendices}
The obtained beamforming vector $\mathbf{v}_c$ exploiting MRT can be expressed as
\begin{align}
	\mathbf{v}_c = (\mathbf{p}\cdot\mathbf{q}_u)^H = \mathbf{p}^H\cdot (\mathbf{f}_u^{H}\mathbf{\Phi}\mathbf{G}+\mathbf{h}^{T}_u)^H,
\end{align}
where $|\mathbf{p}|=N_a$ is obtained according to the Algorithm \ref{alg:alg2}. Given $	\mathbf{v}_c$ and $t_u$, the useful signal $y_c$ received at Bob can be denoted as
\begin{align}\label{29}
y_c &= 	\mathbf{q}_u\tfrac{\mathbf{v}_ce^{-j\mathbf{q}_u\mathbf{v}_{c}}}{||\mathbf{q}_u\mathbf{v}_c||}t_us=\mathbf{q}_u\hat{\mathbf{v}}_c=t_us,
\end{align}
where $\hat{\mathbf{v}}_c = \tfrac{\mathbf{v}_ce^{-j\mathbf{q}_u\mathbf{v}_{c}}}{||\mathbf{q}_u\mathbf{v}_c||}t_us$. Then, the beamforming vector corresponding to $z$ can be represented as
\begin{align}\label{30}
	\hat{\mathbf{v}}_a &= 	\mathbf{w}-\hat{\mathbf{v}}_c.
\end{align}
It can be proven that \eqref{30} is the zero space vector of Bob, and satisfies $\mathbf{q}_e(\hat{\mathbf{v}}_a+\hat{\mathbf{v}}_c) = t_es_es^H$.
\end{appendices}}

\renewcommand\refname{References}
\bibliographystyle{IEEEtran}
\bibliography{mybib}



	
	
	

\end{document}